\documentclass[graybox]{svmult}

\usepackage{amsfonts}
\usepackage{mathrsfs}
\usepackage{amsmath}
\usepackage{amssymb}
\usepackage{dsfont}
\usepackage{bbm}
\usepackage{bm}
\renewcommand{\vec}[1]{\bm{#1}}

\usepackage{mathptmx}       
\usepackage{helvet}         
\usepackage{courier}        
\usepackage{type1cm}        
\usepackage{makeidx}         
\usepackage{graphicx}        
\usepackage{multicol}        
\usepackage[bottom]{footmisc}

\makeindex             

\begin{document}

\title*{The Gauss Law: A Tale\thanks{We dedicate this article to our friend Alberto Ibort on the occasion of his sixtieth birthday.}}
\author{A.P. Balachandran and A.F. Reyes-Lega}
\institute{A.P. Balachandran \at Department of Physics, Syracuse University, Syracuse, New York 13244-1130, USA,\newline \email{balachandran38@gmail.com}
\and A.F. Reyes-Lega \at Departamento de F\'{i}sica, Universidad de los Andes,  A.A. 4976-12340, Bogot\'a, Colombia, \newline\email{anreyes@uniandes.edu.co}}
%

%
\maketitle

\abstract*{The Gauss law plays a basic role in gauge theories, enforcing gauge invariance and creating edge states and superselection sectors. This article  surveys these aspects of the Gauss law in QED, QCD and nonlinear $G/H$ models.
It is argued that nonabelian superselection rules are spontaneously broken.
That is the case with $SU(3)$ of colour which is spontaneously broken to
$U(1)\times U(1)$. Nonlinear $G/H$ models are reformulated as gauge theories and the existence of edge states and superselection sectors in these models is also  established.}

\abstract{The Gauss law plays a basic role in gauge theories, enforcing gauge invariance and creating edge states and superselection sectors. This article  surveys these aspects of the Gauss law in QED, QCD and nonlinear $G/H$ models.
It is argued that nonabelian superselection rules are spontaneously broken.
That is the case with $SU(3)$ of colour which is spontaneously broken to
$U(1)\times U(1)$. Nonlinear $G/H$ models are reformulated as gauge theories and the existence of edge states and superselection sectors in these models is also  established.}

\section{Introduction}
This talk first discusses how locality enters the treatment of the Gauss law constraint and the gauge transformations of its generators,
focussing on the Hamiltonian formalism.
The important role of test functions in a proper treatment of Gauss law becomes apparent.
Choices of various classes of test functions lead to various gauge groups of which the Gauss law-generated gauge group is an invariant subgroup.
There are also gauge transformations of importance (at times loosely called ``large gauge transformations'')
which are not connected to identity.
Observables are local.
A consequence is that they commute with all these gauge transformations.
That leads naturally to a discussion of superselection rules and anomalies:
the latter are just transformations which change the superselection sector and hence are spontaneously broken.
These ideas are illustrated by examples such as axial anomalies and axial flavour transformations of the Standard Model.

We work  with Minkowski spacetime $M^4\cong\mathbb{R}^3\times\mathbb{R}$, where $\mathbb{R}^3$ is the spatial slice.
We also work in the gauge $A_0=0$. The gauge group is defined on the spatial slice $\mathbb{R}^3$.
\section{The Structure of the Gauge Group: the Gauss Law}
Let $H$ be the group which is to be gauged.
In QED, $H$ is $U(1)$; in QCD, it is $SU(3)$.

The gauge group $G$ is not $H$.
Rather, its elements are maps $g$ from $\mathbb{R}^3$ to $H$.
If $g,h\in G$, the multiplication law is ``point-wise'': $gh$ is defined by $(gh)(x)=g(x)h(x)$, $x\in\mathbb{R}^3$.

The Lie algebra $\mathcal{G}$ of the gauge group is associated with the Gauss law in quantum theory:
the latter in Dirac's approach is a condition on state vectors $|\cdot\rangle$:
\begin{eqnarray}
\label{gausslaw}
(D_iE^i+J_0)|\cdot\rangle=0.
\end{eqnarray}
Here, $E^i$ is the electric field, $D_i$ the covariant derivative and $J_0$ is the charge density of the matter field.
If $\lambda_\alpha$ is a basis of generators of the Lie algebra $\underline{H}$ of $H$, we can write
\begin{eqnarray}
 D_iE^i=\partial_iE^i+i A_i^\alpha E^{i,\beta}[\lambda_\alpha,\lambda_\beta], \quad
 E^i=E^{i,\alpha}\lambda_\alpha, \quad A_i=A_i^\alpha\lambda_\alpha, \quad J_0=J_0^\alpha\lambda_\alpha,
\end{eqnarray}
where $A_i^\alpha,E^{j,\beta}$ are canonically conjugate: at equal times,
\begin{eqnarray}
\label{AEconj}
[A_i^\alpha(x),E^{j,\beta}(y)]=i\delta_i^j\delta^3(x-y)\mathbb{I}.
\end{eqnarray}
These expressions are also valid for QED ($H=U(1)$).

The RHS of (\ref{AEconj}) is a distribution.
Therefore, $A_i^\alpha,E^{j,\beta}$ are operator-valued distributions.
Derivatives of distributions are defined by smearing them with test functions and transferring derivatives to test functions.
It turns out to be important to do so for the Gauss law (\ref{gausslaw}).

Thus, let $S_0^\infty(\mathbb{R}^3)$ define $\underline{H}$-valued test functions
$\Lambda=\Lambda^\alpha\lambda_\alpha$ on $\mathbb{R}^3$, $\Lambda^\alpha$ being real, infinitely differentiable and of
fast decrease at infinity. (We will not be precise on the rate of decrease. It is to be adapted to the problem at
hand.) We then write (\ref{gausslaw}) as
\begin{eqnarray}
\label{gausssmeared}
\textrm{Tr}\,\int\left[-(D_i\Lambda)E^i+\Lambda J_0\right]|\cdot\rangle=0.
\end{eqnarray}
If $\textrm{Tr}\,\lambda_\alpha\lambda_\beta=2\delta_{\alpha\beta}$, we can also write (\ref{gausslaw}) as
\begin{eqnarray}
\int\left[-(D_i\Lambda)^\alpha E^{i,\alpha}+\Lambda^\alpha J_0^\alpha\right]|\cdot\rangle=0.
\end{eqnarray}

In the Gauss law (\ref{gausssmeared}), it is important to choose $\Lambda^\alpha$ to vanish at infinity.
It is only then that we can recover the classical Gauss law $D_iE^i+J_0=0$ by partial integration from (\ref{gausssmeared})
without generating surface terms.

\section{The Group $\mathcal{G}_0^\infty$}

The Gauss law generates infinitesimal gauge transformations dependent on $x$.
We can see this as follows.
Let
\begin{eqnarray}
[\lambda_\alpha,\lambda_\beta]=2ic_{\alpha\beta}^\gamma\lambda_\gamma.
\end{eqnarray}
Then,
\begin{eqnarray}
D_iE^i=\partial_iE^{i,\alpha}\lambda_\alpha+2iA_i^\alpha E^{i,\beta}c_{\alpha\beta}^\gamma\lambda_\gamma, \quad\quad
D_iE^{i,\gamma}=\partial_iE^{i,\gamma}-2A_i^\alpha E^{i,\beta}c_{\alpha\beta}^\gamma,
\end{eqnarray}
so that, for example,
\begin{eqnarray}
[D_iE^{i,\gamma}(x),E^{j,\rho}(y)]=-2iE^{j,\beta}(x)c_{\rho\beta}^\gamma\delta^3(x-y).
\end{eqnarray}
This gives, for the smeared Gauss law,
\begin{eqnarray}
\left[\int\left[-(D_i\Lambda)^\alpha E^{i,\alpha}+\Lambda^\alpha J_0^\alpha\right],E^{j,\beta}(y)\right]=2i\Lambda^\gamma(y)c_{\rho\beta}^\gamma
E^{j,\rho}(y),
\end{eqnarray}
which is an infinitesimal $y$-dependent action of $H$.

Now, $\Lambda^\alpha(x)\rightarrow0$ as $|x|\rightarrow\infty$, so that the Gauss law acts trivially at infinity.
Hence, the group $\mathcal{G}_0^\infty$ it generates on exponentiation acts as identity at infinity
(as indicated by the superscript $\infty$).
The elements of $\mathcal{G}_0^\infty$ are also connected to identity as the subscript $0$ indicates,
since they are obtained by exponentiating a Lie algebra element.

If $U(h)$ is the operator representing $h\in\mathcal{G}_0^\infty$, we conclude that on physical states $|\cdot\rangle$,
\begin{eqnarray}
U(h)|\cdot\rangle=|\cdot\rangle.
\end{eqnarray}

\section{The Group $\mathcal{G}_0$: the Emergence of Global Groups}

Let us denote the smeared Gauss law operator as $G(\Lambda)$:
\begin{eqnarray}
G(\Lambda)=\int\textrm{Tr}\,\left[-(D_i\Lambda)E^i+\Lambda J_0\right], \quad\quad \Lambda\in S_0^\infty, \quad\quad
G(\Lambda)|\cdot\rangle=0.
\end{eqnarray}

We now consider more general operators $Q(\mu)$, where the test functions are not required to vanish at infinity:
\begin{eqnarray}
Q(\mu)=\int\textrm{Tr}\,\left[-(D_i\mu)E^i+\mu J_0\right], \quad\quad \mu\in\mathcal{C}^\infty.
\end{eqnarray}
If
\begin{eqnarray}
\mu(x)\big|_{|x|\rightarrow \infty}=\mu\in\mathbb{R},
\end{eqnarray}
there is no reason for $Q(\mu)$ to vanish on physical states.
We call the group that the $Q(\mu)$'s generate as $\mathcal{G}_0$.

Now, $\mathcal{G}_0^\infty$ is a normal subgroup og $\mathcal{G}_0$.
That follows from
\begin{eqnarray}
[\mu,\Lambda]=2i\mu^\alpha\Lambda^\beta c_{\alpha\beta}^\gamma\lambda_\gamma\in S_0^\infty
\end{eqnarray}
for $\mu=\mu^\alpha\lambda_\alpha,\Lambda=\Lambda^\beta\lambda_\beta$, since the RHS tends to zero as $|x|\rightarrow\infty$.

Consider the quotient group
\begin{eqnarray}
\hat{H}=\mathcal{G}_0/\mathcal{G}_0^\infty.
\end{eqnarray}
It is a group that acts non-trivially on quantum states.

Let us assume that $(\mu-\mu_\infty)\in S_0^\infty$, that is, that $\mu$ approaches its asymptotic value rapidly.
Then, if $g(\mu)\in\mathcal{G}_0$, $U(g(\mu))|\cdot\rangle$ depends only on $\mu_\infty$.
That is because if $\mu_{1,\infty}=\mu_{2,\infty}$, $(\mu_1-\mu_2)\in S_0^\infty$ and $G(\mu_1-\mu_2)|\cdot\rangle=0$.

The group $\hat{H}$ is in fact isomorphic to $H$ in simple cases like QED or QCD.
In these cases, we can choose $\mu(x)=\mu_\infty$ for all $x$ and then
\begin{eqnarray}
\label{qmu}
Q(\mu)=\int\textrm{Tr}\,\mu_\infty\left(-A_i^\alpha E^{i,\beta}[\lambda_\alpha,\lambda_\beta]+J_0\right)
\end{eqnarray}
is the familiar $H$-generator after a normalisation of $\mu_\infty$, which for QED is $\mu_\infty=1$.
For QCD, we choose eight $\mu_\infty$'s, $\mu_{\infty,\beta}^\alpha\lambda_\alpha$ ($\beta=1,2,\ldots,8$),
where $\mu_{\infty,\beta}^\alpha=\delta_\beta^\alpha$.
Equation (\ref{qmu}) shows that non-abelian gluons carry non-abelian charges, as the first term in (\ref{qmu}) is not zero.
In QED, instead, the first term is zero, photons having no charge.

But $\hat{H}$ can differ from $H$.
A good example is the 't Hooft-Polyakov model, where $H=U(1)$ while, as Witten has shown~\cite{Witten1979}, $\hat{H}=\mathbb{R}$ in the presence
of magnetic monopoles.
That leads to fractional charges for dyons.

\section{The Sky Group $\hat{\mathcal{G}}_0$}

We can now go one step further and consider the boundary condition
\begin{eqnarray}
\mu(\vec{x})=\mu(|\vec{x}|\hat{x})\xrightarrow[|\vec{x}|\rightarrow\infty]{}\hat{\mu}(\hat{x}).
\end{eqnarray}
That is, allow $\mu$ to approach an angle-dependent limit at infinity.
(We are ``blowing up'' infinity.)
Then, we get the ``Sky'' group $\hat{\mathcal{G}}_0$~\cite{Balachandran2013c,Balachandran1991} with generators
\begin{eqnarray}
\label{rel}
\hat{Q}(\hat{\mu})=\int_{S_\infty^2}\textrm{Tr}\,\left[-(D_i\hat{\mu})E^i+\hat{\mu}J_0\right].
\end{eqnarray}
This group is of importance for discussing infrared effects~\cite{Balachandran2013c}.

\section{Winding Number Gauge Transformations}

These are gauge transformations $h$ which approach $\mathbb{I}$ at infinity.
Hence, they can be regarded as maps from $S^3$ to $H$.
If $h$ has winding number other than zero, then $h$ is a winding number gauge transformation.
We refer to~\cite{Balachandran1991} for the definition and properties of such maps $h$.

If $H$ is $U(1)$ as in QED, then there is no $h$ with non-zero winding number: $\Pi_3(U(1)_g)=0$.

If $H$ is a compact, simple Lie group like $SU(N)$, there are such transformations.
Let $h_1$ be one such typical transformation with winding number one.
Then, powers of $h_1$, namely $h_1^k$, $k\in\mathbb{Z}$, generate the group $\mathbb{Z}$.

If $\mathcal{G}_W^\infty$ is the gauge group with elements becoming identity at infinity, but not necessarily connected to identity,
then the winding number group is $\mathcal{G}_W^\infty/\mathcal{G}_0^\infty$.

We can relax the condition at infinity and consider $\hat{\mathcal{G}}_W$.
With $\hat{\mathcal{G}}_W$, we allow angle-dependence at infinity.
This group may not be connected to identity.
Then again we have that $\hat{\mathcal{G}}_W/\hat{\mathcal{G}}_{0}$
(the subscript $0$ as usual denoting the component connected to identity) is the winding number group:
\begin{eqnarray}
\hat{\mathcal{G}}_W/\hat{\mathcal{G}}_{0}\approx\mathcal{G}_W/\mathcal{G}_0.
\end{eqnarray}
The winding number group (``deck transformations'') is responsible for the $\theta$-vacua of QCD.

An important point is that in QCD and $SU(N)$ gauge theories, we cannot write winding number operators in terms of field variables like $A$
and $E$.
(An exception occurs in the 't Hooft-Polyakov model mentioned above.)
Still, they are well-defined as automorphisms of local observables.
There may be cases where they are not implementable as unitary operators, leading to their ``spontaneous'' breakdown.

We now have a list of various gauge groups with their mutual relations:
\begin{eqnarray}
\begin{array}{cccccccccc}
&\hat{\mathcal{G}}_W & \supset & \mathcal{G}_W & \supset & \mathcal{G}_0 & \supset & \mathcal{G}_0^\infty  =
\textrm{Gauss law group}\\
&\cup & {} & \cup & {} & \cup & {} & {} \\
&\hat{\mathcal{G}}_W^\infty & = & \mathcal{G}_W^\infty & \supset & \mathcal{G}_0^\infty & {} & {}
\end{array}
\end{eqnarray}
In each non-trivial inclusion here, the subgroup is normal.
Only the Gauss law group necessarily acts as identity on quantum states.
Also, groups with index $W$ do not have operators given by the canonical approach in QCD.

\section{On Local Observables and Gauge Invariance}

Let $\mathcal{A}$ be the algebra of local observables.
If $\varphi$ is a local quantum field of the Lagrangian approach and $f$ is a test function with compact support $K$, then
\begin{eqnarray}
\label{varphif}
\varphi(f)=\int d^3x\, f(x)\varphi(x)
\end{eqnarray}
or, better, $e^{i\varphi(f)}$ is an element of $\mathcal{A}$.
Here, for illustration, we assume that $\varphi$ is a scalar.
We can exponentiate $\varphi(f)$ in (\ref{varphif}) to get a unitary (and hence bounded) operator.

In our approach, $K$ is a spatial region.
A more rigorous formulation will require $K$ to be a spacetime region~\cite{Haag1996}.

If $a\in\mathcal{A}$, thought of as an operator in a Hilbert space $\mathcal{H}$ of physical states, then we have that
\begin{eqnarray}
\label{comm}
aU(h)=U(h)a \quad\quad \textrm{if } h\in\mathcal{G}_0^\infty.
\end{eqnarray}
That is because we want both $|\cdot\rangle$ and $a|\cdot\rangle$ to be in the kernel of the Gauss law.
Thus,
\begin{eqnarray}
[a,G(\Lambda)]=0 \quad\quad \textrm{if } \Lambda\in S_0^\infty.
\end{eqnarray}
The result (\ref{comm}) follows from here on exponentiation of $G(\Lambda)$.

But $a$ is local.
An important consequence is then that $a$ commutes with all the $\mathcal{G}$-groups.

Let $\mathcal{A}^\prime$ be the commutant of $\mathcal{A}$ and $\mathbb{C}\mathcal{G}_W$ the group algebra of $\mathcal{G}_W$.
Then, the above claim means the following: because of locality and the Gauss law constraint, we have that
\begin{eqnarray}
\mathcal{A}^\prime\supseteq\mathbb{C}\mathcal{G}_W.
\end{eqnarray}

It is enough to show the infinitesimal version of this result for all $\mathcal{G}$'s except $\mathcal{\hat G}_W$'s
(Towards the end of this section, we consider $\mathbb{C}\mathcal{G}_W$).
For the former, there is the generator $Q(\mu)$.
Let $\varphi(f)$ be a local field supported in a compact region $K$.
The commutator $[Q(\mu),\varphi(f)]$ depends only on $\mu\big|_K$, the restriction of $\mu$ to $K$, because of locality.
That is, if $\mu\big|_K$ and $\nu\big|_K$ are equal, then
\begin{eqnarray}
[Q(\mu)-Q(\nu),\varphi(f)]=0.
\end{eqnarray}
So let us extend $\mu\big|_K$ to a $\nu\in C_0^\infty$ in any manner with the only condition $\nu\big|_K=\mu\big|_K$.
Then,
\begin{eqnarray}
[Q(\mu),\varphi(f)]=[Q(\nu),\varphi(f)]=[G(\nu),\varphi(f)]=0.
\end{eqnarray}
The second equality holds true because $\mu,\nu\in C_0^\infty$.
This proves the result.

The proof for winding number transformation is along the same lines.
If $g\in\hat{\mathcal{G}}_W$, then $U(g)\varphi(f)U^{-1}(g)$ depends only on $g\big|_K$ and $g^{-1}\big|_K$.
We now extend $g^{-1}\big|_K$ outside $K$, so that it has globally zero winding number and belongs to $\mathcal{G}_0^\infty$.
If the extended gauge transformation is $h$, then $U(h)\in\mathcal{A}^\prime$.
Hence, $U(g)\in\mathcal{A}^\prime$ too.
We can thus conclude that $\hat{\mathcal{G}}_W\in\mathcal{A}^\prime$.

\section{On Superselection Groups}

These are the transformations commuting with $\mathcal{A}$.
Hence, the group $\hat{\mathcal{G}}_W$ is associated with the superselection group.

The subgroup $\mathcal{G}_0^\infty$ of Gauss law becomes $\mathbb{I}$ on quantum states.
It is normal in $\hat{\mathcal{G}}_W$.
Hence, it is more appropriate to identify $\hat{\mathcal{G}}_W/\mathcal{G}_0^\infty$ or a subgroup thereof with the superselection group.

The group algebra $\mathbb{C}(\hat{\mathcal{G}}_W/\mathcal{G}_0^\infty)$ commutes with $\mathcal{A}$:
$\mathbb{C}(\hat{\mathcal{G}}_W/\mathcal{G}_0^\infty)\in\mathcal{A}$,
it is a Hopf algebra.
For more discussions on Hopf algebras, see~\cite{Balachandran2010}.
We can also work with $\mathbb{C}(\hat{\mathcal{G}}_W/\mathcal{G}_0^\infty)$ to illustrate superselection theory.

Subgroups of $\hat{\mathcal{G}}_W/\mathcal{G}_0^\infty$ give us the familiar superselection rules of QED and QCD.
That is because of the following.
Local observables cannot change the irreducible representation (IRR) $\rho$ of $\hat{\mathcal{G}}_W/\mathcal{G}_0^\infty$ on the
Hilbert space $\mathcal{H}_\rho$
(we now label $\mathcal{H}$ also with $\rho$): $\mathcal{A}\mathcal{H}_\rho\subseteq\mathcal{H}_\rho$.
Hence, by definition, $\hat{\mathcal{G}}_W/\mathcal{G}_0^\infty$ is superselected.

More will be said below when the group in question is non-abelian.
We now illustrate these remarks.

\subsection{Charge and Colour}

In QED,  ${\mathcal{G}}_0/\mathcal{G}_0^\infty=U(1)$.
Thus, charge is superselected.
So is colour, since  ${\mathcal{G}}_0/\mathcal{G}_0^\infty=SU(3)_c$ in QCD.

There is a subtlety about colour because it is non-abelian.
In a unitary IRR (UIRR), as Dirac has explained, we can diagonalise a maximal commuting set (CCS)
of operators from its group algebra
$\mathbb{C}SU(3)$.
A choice for the CCS is $C_2,C_3,I,I_3,Y$, where $C_2,C_3$ are the quadratic and cubic Casimir operators, $I$ is isospin
and $I_3,Y$ are the operators representing $\lambda_3/2$ and $\lambda_8/\sqrt{3}$ of the Gell-Mann matrices.
Thus, a vector state in a UIRR $\rho$ is characterised by the eigenvalues $c_2,c_3,i,i_3,y$ of these operators:
\begin{eqnarray}
\mathcal{H}_\rho: \quad |c_2,c_3,i,i_3,y,\cdot\rangle.
\end{eqnarray}
No operator of $\mathcal{A}$ will change these eigenvalues.
But a generic element of $SU(3)_c$
(for example, $U_\alpha$ and $V_\alpha$, the $U-$ and $V-$ spin operators of~\cite{Lipkin2002}) will change $i_3,y$.
Hence, they cannot be implemented on $\mathcal{H}_\rho$.\\
Operators changing $\mathcal{H}_\rho$ are said to be \textit{spontaneously broken} or \textit{anomalous}.
Hence, in QCD, $SU(3)_c$ of colour is spontaneously broken to $S\left(U(1)_{I_3}\times U(1)_Y\right)$, generated by $I_3,Y$~\cite{Balachandran2013c}.
Analogous results have been found for the ethylene molecule~\cite{Balachandran2013d} and for colour breaking by the non-abelian monopoles of
grand unified theories \cite{Balachandran1983a,Zaccaria1983,Balachandran1984,Balachandran1984a}.

{\bf Remark:}
We make no distinction between spontaneous symmetry breaking and symmetry breaking by anomalies.
Both are of the same origin: they change the domain of the observables $\mathcal{A}$.
In the above, anomalous operators change $\mathcal{H}_\rho$.
(It was Esteve who first discussed anomalies as transformations changing the domain of the Hamiltonian~\cite{Esteve1986}.)

\subsection{QCD $\theta$-vacua}

These originate in $\mathcal{G}_W$.
It is enough to consider $\mathcal{G}_W^\infty$.
Let $T$ be a winding number one transformation.
Then,
\begin{eqnarray}
U(T)a=aU(T) \quad\quad \textrm{if } a\in\mathcal{A}
\end{eqnarray}
because of locality.
It also preserves the Gauss law constraint: if $G(\Lambda)|\cdot\rangle=0$, then $G(\Lambda)U(T)|\cdot\rangle=0$.
Hence, in an IRR of $\mathcal{A}$, we can diagonalise $U(T)$.

For $H=SU(2)$, a typical winding number one transformation is
\begin{eqnarray}
T(\vec{x})=\cos\theta(r)+i\vec{\tau}\cdot\hat{x}\sin\theta(r), \quad r=|\vec{x}|, \quad \theta(0)=-\pi, \quad \theta(\infty)=0.
\end{eqnarray}

If $T$ has winding number one, $T^k$ has winding number $k$. So the group it generates is $\mathbb{Z}$.
The UIRR's of $\mathbb{Z}$ are labelled by the points on a circle $S^1=\{e^{i\theta}\}$.
If the quantum state is characterised by $\rho_\theta: \, T\rightarrow U(T)=e^{i\theta}$, we have that
\begin{eqnarray}
U(T)|e^{i\theta},\ldots\rangle=e^{i\theta}|e^{i\theta},\ldots\rangle.
\end{eqnarray}
These are the $\theta$-states of QCD.
There is an extensive literature on $\theta$-states.

\subsection{The Sky Group}

This group emerged from the study of infrared problems in QED.
The name was suggested by the Scri or BMS group of Bondi, Metzner and Sachs~\cite{Penrose1974,McCarthy1972} and the Spi group of Ashtekar~\cite{Ashtekar1992}.

The Sky group $\mathcal{G}$ has generators $Q(\mu)$, where
\begin{eqnarray}
\mu(\vec{x})=\mu(r\hat{n})\xrightarrow[r\rightarrow\infty]{}\mu(\hat{n}),
\end{eqnarray}
and where $\mu(\hat{n})$ need not be zero.

There is an operator, an intertwiner, $V(\omega)$  which maps state vectors with $Q(\mu)=0$ to ones where it is not zero.
Then, there are the sectors with ``in" state vectors (cf.~\cite{Balachandran2016a,Balachandran2018} and references therein)
\begin{eqnarray}
\label{reprA}
&& e^{q_n\int d^3x [A_i^-(x)\omega_i^+(x)-A_i^+(x)\omega_i^-(x)]}|n, P,\cdot\rangle:=|n, P, \omega,\cdot\rangle, \\
&& |n, P, 0,\cdot\rangle\equiv|n, P, \cdot\rangle, \quad\quad q_n\neq 0
\end{eqnarray}
created by the infrared photons.  Here $A_i^\pm$ are the positive and negative frequency parts of the electromagnetic potential in the Coulomb gauge, and the functions $\omega_i^+$, $\omega_i^- = \bar{\omega}_i^+$  are transverse:
\begin{eqnarray}
\partial_i\omega_i^\pm(x)=0,
\end{eqnarray}

Also they do not vanish fast as we approach infinity:
\begin{eqnarray}
\lim_{r\rightarrow\infty} r^2\,\hat{x}_i\,\omega_i(x)^\pm\neq 0.
\end{eqnarray}
One such typical $\omega_i^+$ has the Fourier transform
\begin{eqnarray}
\label{omega}
\hat{\omega}_i^+(k)=\int d^3x\, e^{i\vec{k}\cdot\vec{x}}\omega_i^+(x)=\frac{1}{P\cdot k+i \epsilon}
(P_i-\vec{P}\cdot\hat{k} \,\hat{k}_i)
\end{eqnarray}
(with  $\epsilon$ decreasing to zero as usual).
The momentum $P_\mu$ is the total momentum of the charged system. (We have not shown the individual momenta and charges
of which $P$ and $q_n$ are composed as they are not importatnt for our considerations.) The important point here is that
$\hat{\omega}_i^+$ is
not square-integrable:
\begin{eqnarray}
\langle \omega,\omega\rangle:=\lim_{|\vec{k}'|\rightarrow 0}\int^\infty_{|\vec{k}'|}
\frac{d^3k}{2|\vec{k}|}|\hat{\omega}_i^+(k)|^2=\infty.
\end{eqnarray}
It is then a theorem~\cite{Roepstorff1970} that the representation of $\mathcal{A}$ built on (\ref{reprA})
is superselected: it is not the Fock space representation.

$V(\omega)$ commutes with the Gauss law operator $G(\Lambda)$.
But that is not the case with $Q(\mu)$:
\begin{eqnarray}
\label{relat}
e^{iQ(\mu)}V(\omega)=\exp\left[q_n\lim_{r^2\rightarrow\infty}\int_{S_\infty^2}d\Omega\, r^2\mu(\hat{n})\hat{n}_i(\omega_i^+-\omega_i^-)
(r\hat{n})\right]
V(\omega)e^{iQ(\mu)},
\end{eqnarray}
where $S_\infty^2$ is the ``sphere'' at infinity (we are ``blowing up'' infinity).

The algebra defined by the relation (\ref{relat}) is useful to study the infrared effects in gauge theories and their phenomenology~\cite{Balachandran2015}.

There are non-abelian and gravitational generalisations of $V(\omega)$~\cite{Balachandran2015}

If $|\cdot\rangle$ is a vector in the Fock space, then $V(\omega)|\cdot\rangle$ is not in the Fock space since $\omega_i$ is not square-integrable:
\begin{eqnarray}
\int d^3x\,|\omega_i(x)|^2=\infty.
\end{eqnarray}
Using this fact, one proves that  Lorentz invariance is broken: boost operators cannot be defined on $V(\omega)|\cdot\rangle$.
Colour and $SL(2,\mathbb{C})$ gauge symmetry are similarly broken.
The discussion of these issues may be found in papers.

\section{Global Symmetries: Lorentz and Flavour Groups}

Apart from gauge groups, whose elements are spacetime dependent, we have in addition global symmetries.
They transform \textit{all} local fields and cannot be localised in a compact region $K$.
A simple example is spatial translation.
For a free scalar field, its generators are
\begin{eqnarray}
P_i=\frac{1}{2}\int d^3\vec x\,\left[\varphi(x),\partial_i\Pi(x)\right], \quad\quad
[\varphi(x),\Pi(y)]\Big|_{x^0=y^0}=i\delta^3(\vec x-\vec y),
\end{eqnarray}
as deduced from the standard Lagrangian.
They involve a density such as $\varphi(x)\partial_i\Pi(x)$ integrated over \textit{all} space.
For these reasons, they are not local.
We call them \textit{global}.

On local observables $\mathcal{A}$, global symmetries act as automorphisms.
Unitary elements of $\mathcal{A}$ also act as automorphisms of $\mathcal{A}$.
The latter generate the \textit{inner} automorphism group $\textrm{Inn}\,\,\mathcal{A}$.
If $\textrm{Aut}\,\,\mathcal{A}$ is the group of all automorphisms of $\mathcal{A}$, then $\textrm{Inn}\,\,\mathcal{A}$ is a normal subgroup of
$\textrm{Aut}\,\,\mathcal{A}$.
Global symmetries are elements of the quotient group $\textrm{Aut}\,\,\mathcal{A}/\textrm{Inn}\,\,\mathcal{A}$, which is called the outer
automorphism group $\textrm{Out}\,\,\mathcal{A}$.

There is no guarantee that global symmetries, that is $\textrm{Out}\,\,\mathcal{A}$, can be implemented by operators in an IRR $\rho$ of
$\mathcal{A}$ on $\mathcal{H}_\rho$.
Superselection operators are multiples of identity on $\mathcal{H}_\rho$ and it can happen that elements of $\textrm{Out}\,\,\mathcal{A}$ change
$\rho$.
In that case, they are spontaneously broken.
We know many such examples.
We list a few below.

\subsection{Axial $U(1)$ Anomaly}

In quantum physics, we seek a representation of $\mathcal{A}$ which preserves the domain $\mathcal{D}_H$ of the Hamiltonian:
$\mathcal{A}\mathcal{D}_H\subseteq\mathcal{D}_H$.
This is important so that we have well-defined time evolution.

But, as Esteve discusses~\cite{Esteve1986}, axial $U(1)$ transformations $U(1)_A$ change $\mathcal{D}_H$, that is, the IRR of $\mathcal{A}$.
Hence, they are spontaneously broken.
That is so in QED and the Standard Model.

\subsection{The Axial Flavour Anomaly}

The flavour group in QCD at the Lagrangian level is $U(N_f)_L\times U(N_f)_R$ (up to discrete groups) acting on the
left- and right-handed quarks. If $(g,h)$ is a transformation of this group, they are interchanged by parity $P$:
\begin{eqnarray}
P: \,\, (g,h)\rightarrow (h,g).
\end{eqnarray}
Vector transformations $(g,g)$ commute with $P$.
The axial transformations $(g,g^{-1})$ do not.
The latter are all anomalous.

\subsection{How QED Breaks Lorentz Invariance}

A very striking example occurs in QED, where, as proved by
Buchholz\footnote{In \cite{Buchholz1986}, Buchholz has proven that Lorentz transformations must be
spontaneously broken in electrically
charged sectors and also that  electrically charged states cannot be eigenstates of the mass
operators.}~\cite{Buchholz1986} and
Fr\"ohlich, Morchio and Strocchi~\cite{Froehlich1979,Froehlich1979a}, infrared effects dress the charged particle
states by a $V(\omega)$ as in~\cite{Froehlich1979,Froehlich1979a}, $\omega$ being known, and change the Fock space to a
non-Fock space. In this new representation space, boost generators and hence the Lorentz group are spontaneously
broken\footnote{Fr\"ohlich, Morchio and Strocchi base their work on the asymptotic
fields (relying on Buchholz's collision theory for massless Bosons~\cite{Buchholz1977}).
In ~\cite{Buchholz1982}, Buchholz relates the problem
to the interacting (time zero) fields, i.e. Gauss law.}.
For an important generalization of these works, see~\cite{Buchholz2014}.

There are extensions of this result to QCD~\cite{Balachandran2016a}.

\subsection{The Higgs Field}

The Higgs field is of standard use for the spontaneous breaking of symmetry.
We discuss it briefly in the context of the group $U(1)$, such as in superconductivity.

So let $\phi$ be a charged Higgs field approaching the constant value $\phi_\infty$ at spatial infinity.
If $a$ is a local observable, then
\begin{eqnarray}
\lim_{r\rightarrow\infty}[a,\phi(x)]=0, \quad\quad r=|\vec{x}|,
\end{eqnarray}
so that $\phi_\infty$ is superselected.

We consider $U(1)$ as a gauge symmetry.
So, if $u\in U(1)$, then
\begin{eqnarray}
U(u)a=aU(u)
\end{eqnarray}
and $U(u)$ too commutes with all local observables.

But
\begin{eqnarray}
U^{-1}(u)\phi(x)U(u)=u\phi(x),
\end{eqnarray}
so that $U(u)$ and $\phi_\infty$ do not commute.
These superselected operators form a \textit{non-abelian} group.
Each superselected sector, as discussed earlier, can be labelled by the eigenvalues of only \textit{one} of them.

But we want to preserve the domain $\mathcal{D}_H$ of the Hamiltonian $H$.
The latter has a potential $V(\phi)$ which is zero only at $\phi_\infty$.
That means that in $\mathcal{D}_H$, the operator for $\phi(x)$ must approach $\phi_\infty$ as $r\rightarrow\infty$.
Hence, we label the superselection sector by $\phi_\infty$.
But then $U(u)$ changes $\phi_\infty$ to $u\phi_\infty$ and is spontaneously broken.

It is possible to express $\phi_\infty$ in terms of the field $\phi$ smeared with a test function and its expectation value for vectors
$|\cdot\rangle\in\mathcal{D}_H$.
\section{Non-linear Models and Edge Excitations}
Consider a model for Goldstone modes with gauge group $G$ which is spontaneously broken to $H\subset G$. Then the model describes Goldstone modes with target space $G/H$. If the model can be described as a gauge theory, then we can apply the previous discussion. This can be done as follows~\cite{Balachandran1979}. We fix an orthonormal basis of the Lie algebra of $G$:
\begin{eqnarray}
T(\alpha),  &\qquad& \alpha=1,2,\ldots, |H|,\\
S(i), &\qquad& \mbox{remaining generators of } G.
\end{eqnarray}
Then under an action of $h\in H$,
\begin{eqnarray}
hT(\alpha)h^{-1} & = & T(\beta) h_{\beta\alpha},\\
hS(i)h^{-1} & = & S(j)D_{ji}(h).
\end{eqnarray}
Set
\begin{eqnarray}
A_\mu (g) & = & T(\alpha) \textrm{Tr } T(\alpha) g^{-1}(x) \partial_\mu g (x),\\
B_\mu (g) & = & S(i)\textrm{Tr } S(i)g^{-1}(x)\partial_\mu g(x).
\end{eqnarray}
Then under the right action of $H$,
\begin{eqnarray}
A_\mu (gh) & =& h^{-1} A_\mu (g) h + h^{-1}\partial_mu h,\\
B_\mu (gh) & =& h^{-1} B_\mu (g) h,
\end{eqnarray}
i.e. $A_\mu$ is a connection while $B_\mu$ is a tensor field.

For gauge group $\mathcal H\ni h:\mathbb R^n\rightarrow H$, we can write Lagrangian densities like
\begin{equation}
\mathcal L_1 = -\lambda \textrm{Tr } B_\mu (g) B^\mu (g),
\end{equation}
or
\begin{equation}
\mathcal L_2 = -\lambda \textrm{Tr } F_{\mu\nu} (A) F^{\mu\nu}(A).
\end{equation}
They reduce to standard $\sigma$-model Lagrangians, e.g. with $G=SU(2)$, $H=U(1)$, so that $G/H=S^2$. Explicitly writing:
\[
g(x)\sigma_3 g^{-1}(x)=\sigma_\alpha \varphi_\alpha(x) \Rightarrow \varphi_\alpha (x)\varphi_\alpha (x) = \mathds 1,
\]
we get
\begin{eqnarray}
\mathcal L_1 &\sim & -\lambda (\partial_\mu \varphi_\alpha)(\partial^\mu \varphi_\alpha),\nonumber\\
\mathcal L_2 &\sim & -\varepsilon^{\alpha\beta\gamma}\lambda (\varphi_\alpha \partial^\mu \varphi_\beta \partial^\nu \varphi_\gamma)^2.
\end{eqnarray}
But (non-local) observables need be invariant only under
\begin{equation}
\mathcal H_\infty = \lbrace h\in\mathcal H\;|\;h_\infty (\hat x) = \lim_{r\rightarrow\infty} h(r\hat x)=\mathds 1\rbrace.
\end{equation}
Can we find such observables invariant only under $\mathcal H_\infty$ and not under $\mathcal H$?
Consider the Wilson line
\begin{equation}
W(g,x,e)= \exp \int_\infty^xd\lambda e^{\mu} A_\mu (g(x+\lambda e)),
\end{equation}
where $e^\mu$ is a spacelike unit vector. Under gauge transformation by $h\in \mathcal H$,
\begin{equation}
W(g,x,e)\rightarrow h_\infty (\hat x) W(g,x,e) h^{-1}(x).
\end{equation}
Hence
\begin{equation}
\tilde B_\mu (g,x,e) \equiv W(g,x,e) B_\mu (x)[W(g,x,e)]^{-1}
\end{equation}
is invariant by small, but not by large gauge transformations. $\tilde B_\mu (g,x,e)$ is \emph{not} a local field. $\tilde B_\mu (g,x,e) |x\rangle$ is a state with edge excitations. How do we see them? Perhaps through instantons. Thus we have the $\theta$-vacuum term
\[
\frac{\theta}{32\pi^2}\int \mbox{Tr} F(A)\wedge F(A)
\]
that we can add to the action. There are also instanton solutions of $F=\ast F$ (for certain groups, the ADHM method works).But this topological term cannot be reduced to an integral of standard $G/H$-model fields. It violates $CP$ invariance and can induce electric dipole moment. The present limit is given by
\[
\theta \leq 10^{-10}.
\]
Some of these ideas extend to self-dual gravity as well.

\begin{acknowledgement}
The second author acknowledges financial support from the Faculty of Sciences of Universidad de los Andes (convocatoria
2018-2019 para la Financiaci\'on de Programas de Investigaci\'on, programa
 ``Geometric, Algebraic and Topological Methods for Quantum Field Theory'').

\end{acknowledgement}
%

%
%
%

\end{document}